\documentclass[conference]{IEEEtran}
\usepackage{comment}
\usepackage{booktabs}
\usepackage{color}
\usepackage{balance}
\usepackage{hyperref}
\usepackage{flushend}
\usepackage{tablefootnote}
\usepackage{cite}
\usepackage{url}
\usepackage{xcolor}
\usepackage{amsmath}
{\begin{center}\vspace{0mm}\noindent\begin{Sbox}\begin{minipage}{0.95\columnwidth}}%
{\end{minipage}\end{Sbox}\fbox{\TheSbox}\end{center}\vspace{0mm}}

\usepackage{color}
\usepackage{ifthen}
\newboolean{showcomments}
\setboolean{showcomments}{true} 
\ifthenelse{\boolean{showcomments}}
  {

        \newcommand{\dmg}[1]{{\color{blue}\emph{dmg says: #1}}\xspace}
        \newcommand{\todo}[1]{\textcolor{red}{{\sc #1}}}
        \newcommand{\internalnote}[1]{\marginpar{\scriptsize note: #1}}
        
		\newcommand{\remarks}[1]{\color{red}[#1]\color{black}}

		\newcommand{\del}[1]{\textcolor{red}{\sout{#1}}} 
  }
  {
        \newcommand{\dmg}[1]{}
        \newcommand{\todo}[1]{}
        \newcommand{\internalnote}[1]{}
		\newcommand{\remarks}[1]{}

		\newcommand{\del}[1]{} 
  }

\ifCLASSINFOpdf
  \usepackage[pdftex]{graphicx}
\else
  \usepackage[dvipdfmx]{graphicx}
\fi
\newcommand{\health}{{Workforce}}
\newcommand{\healthAbbv}{{WF}}
\newcommand{\pullR}{{Pull Request}}
\newcommand{\PR}{{PR}}
\newcommand{\wealth}{{Gross Product Pull Requests}}
\newcommand{\wealthAbbv}{{GPPR}}
\newcommand{\RQone}{{Does Wealth influence how OSS communities maintain their Health?}}
\newcommand{\RQtwo}{{Does Health influence how OSS communities maintain their Wealth?}}
\newcommand{\typeone}{{\textsc{homebrew}}}
\newcommand{\typetwo}{{\textsc{bitcoin}}}
\newcommand{\typethree}{{\textsc{d3}}}
\newcommand{\movieurl}{{goo.gl/Ig6NTR}}
\begin{document}
%
\title{The Health and Wealth of OSS Projects: Evidence from Community Activities and Product Evolution}

\author{\IEEEauthorblockN{Saya Onoue, Raula Gaikovina Kula, Hideaki Hata, Kenichi Matsumoto}
\IEEEauthorblockA{Graduate School of Information Science\\
Nara Institute of Science and Technology, Nara, Japan\\
Email: \{onoue.saya.og0, raula-k, hata, matumoto\}@is.naist.jp}
}


%


\maketitle

\begin{abstract}
\textit{Background:} Understanding the condition of OSS projects is important to analyze features and predict the future of projects.
In the field of demography and economics, health and wealth are considered to understand the condition of a country.
\textbf{Aim:} In this paper, we apply this framework to OSS projects to understand the communities and the evolution of OSS projects from the perspectives of health and wealth.
\textit{Method:} We define two measures of \health\ (\healthAbbv) and \wealth\ (\wealthAbbv). 
We analyze OSS projects in GitHub and investigate three typical cases.
%
\textit{Results:} We find that wealthy projects attract and rely on the casual workforce. 
Less wealthy projects may require additional efforts from their more experienced contributors.
\textit{Conclusions:} This paper presents an approach to assess the relationship between health and wealth of OSS projects. 
An interactive demo of our analysis is available at \url{\movieurl}.
\end{abstract}

%
\IEEEpeerreviewmaketitle

\section{Introduction}



It is well-known that Open Source Software (OSS) components play a critical role of contemporary software development.
With the emergence of open repositories like GitHub, OSS projects is now easily accessible and have been featured as popular and impactful software such as Linux, Ubuntu and Firefox OSS applications.
	
The onion model has been widely studied for sustainable OSS development communities \cite{Nakakoji2002,Kishida2003,Aberdour2007}, depicting an OSS community as being one dimension onion-like shape. Crowston and Howison concluded that assessing the health of an OSS project is not an easy task \cite{Crowston2006}.
Due to the Bazaar-like structure and altruistic nature of contributors, it is hard to assess the success and livelihood of an OSS project. 
Furthermore, defining the success of OSS projects is difficult. 
Senyard et al. studied how a project can establish this community and be a success \cite{Senyard2004}.
Targeting the initial phases of free software projects, they explain what is needed to facilitate the bazaar phase and remain successful.
Many studies have categorized contributors based on their contributions.
Marco et al. \cite{Steinmacher2015} discovered empirical evidence of the barriers faced by newcomers to OSS projects when placing their first contribution.
They say that onboarding is important for online communities to leverage outsider contribution, and conclude that a smooth first contribution may increase the total number of successful contributions made by both single and long-term contributors. 

In this study, we would like to assess OSS projects in the two dimensions of health and wealth.
Our work is inspired by Hans Rosling's talk on Health and Wealth of Nations\footnote{Hans Rosling: The best stats you've ever seen, TED 2006. \url{http://www.ted.com/talks/hans_rosling_shows_the_best_stats_you_ve_ever_seen}}.
The Health and Wealth visualization depicts how life expectancy of nations is correlated to their economics.
Similarly, we would like to understand the Health and Wealth in terms of OSS communities and their projects. 
Concretely, we represent Health as the community activities such as contributor work rate, while Wealth is represented as the product evolution, which can be described as the completed pull requests over time. 
In a preliminary evaluation study, we use three case studies selected from 90 OSS projects to highlight the relationship between the Health and Wealth of an OSS project. 
The following research questions guide our study:
\begin{description}
\item[$RQ_1$:] \RQone
\item[$RQ_2$:] \RQtwo
\end{description}
From the results, we observe the following:
\begin{itemize}
\item \textit{Less wealthy projects may require additional efforts from their more experienced contributors.}
\item \textit{Wealthy projects attract and rely on the inexperienced casual workforce.}
\item \textit{Experienced contributors are major workforce of an OSS project.}
\item \textit{Wealthy projects do not depend only on higher workforce if they have sufficient casual workforce.}
\end{itemize}
Furthermore, more wealthy projects rely on casual contributors to maintain their Health.
We envision that the added dimension of Wealth adds an economic values to the assessment of OSS projects and may lead to use understanding more about the success and livelihood of OSS projects.
An interactive demo that shows the evolution of Health and Wealth of the studied OSS projects is available at \url{\movieurl}.

 
\section{Health and Wealth in OSS Projects}
\subsection{Basic Concepts}
The basic concepts of Health and Wealth are best represented by Tom Carden and Gapminder in their \textit{``The Wealth \& Health of Nations''}\footnote{The Wealth \& Health of Nations \\\url{https://bost.ocks.org/mike/nations/}}.
They discuss Health and Wealth in the context of nations, showing that these two factors are very strongly related. 
Rosling says that 80\% of the variance of Health can be explained by Wealth. 
\textit{``This means that we know that increased Wealth is extremely strongly correlated with longer lifespans. 
There are some interesting details however. 
It seems that you can advance much faster as a nation if you are healthy first than if you are wealthy first''}.
Rosling then further elaborates: \textit{“Health cannot be bought at the supermarket.”} 
He makes it clear that Health is an investment. 
He further explains how \textit{``You have to build infrastructure, you have to train people, and you have to educate the whole population.''}

Our key idea in the paper is to show how these concepts can be utilized in an OSS setting. 
In country economics, the following metrics are considered to be health and wealth: Life Expectancy is a statistical measure of the average time persons are expected to live, while Gross Domestic Product (GDP) measures the total of goods and services produced in a given year within the borders of a given country\cite{capital}. 

\subsection{Health and Wealth Metrics in an OSS setting}
Unlike countries of the world, OSS project are often depicted as Bazaar-like, with no set structure or organization\cite{bird2008,Bird2011,Raymond2012}. 
In this setting, we propose that the two main factors that influence the livelihood of an OSS project: (i) the activities of its community members and (ii) the evolution of the product itself.

We first define the OSS Health as a indicative of three factors of how community activities are performed in a project based on our previous studies: (a) workrate (defined as labor) of each contributor\cite{6754343}, (b) attractiveness of new contributors to a community\cite{Onoue2014}, and (c) active retention of experienced members\cite{Onoue2016}.
We measure labor as community contributions within the projects. 
Similar to Rigby et al.\cite{Rigby2014}, we take into account the contributor experience to evaluate attractiveness and retention factors.
We only consider any source code changes as contributions (i.e., comments made by contributors are ignored). 

Let $Labor_{i,m}$ be the number of contributions an individual $i$ has made in month $m$.
We call this weighted measure of $Labor$ as \textbf{\health~(\healthAbbv)}, where the \healthAbbv\ for a contributor $i$ in month $m$, who has an experience of working $e_i$ months, is formally defined as follows:

\begin{equation*}
\begin{split}
\healthAbbv_{m}(e_i)&=\sum_{j=1}^{e_i} \frac{Labor_{i,j}}{e_i-j+1}
\end{split}
\end{equation*}


Using $e_i$ as the number of months since a contributor $i$ first joined the project, the function {\it $\healthAbbv(m)$} describes the monthly contributions of the member to the OSS project.
We use a linear decay function to consider a factor of the contributor's experience (i.e., more experienced contributor will have less weight than a newcomer to the projects in their recent contributions).

We use the median of all $\healthAbbv_m(e_i)$ of contributors in month $m$, namely, $\widetilde{\healthAbbv_m}$, as the indicator of the OSS Health in month $m$.
By obtaining median, we can see the mediated workforce among highly active contributors and casual contributors.

Our definition of OSS Wealth is based on the evolution of product, accumulated source code patches. 
One approach to measure the evolution, especially for projects that use Git version control system, is by the number of \pullR s (\PR). 
Pull requests tells other contributors about changes that you wish to make to the product.
Once a \PR~is opened, contributors can discuss and review the potential changes with the community and can add follow-up commits before the changes are merged into the product source code.
After the change is merged, the \PR~will be closed.

We define \textbf{\wealth~(\wealthAbbv)} as the number of completed \pullR s in month $m$. 
To add weight on more recent \PR s, we use a weighted measure $\PR _{m}(pr)$ to return the number of months that a \PR~$pr$ took to close (i.e., \PR $_m \geq$1).
Thus, \PR s taking more than a month to complete has less weighting.
\wealthAbbv$_m$ is formally defined as:

\begin{equation*}
\begin{split}
\wealthAbbv_m&= \sum_{pr\in \wealthAbbv} \frac{1}{PR_{m}(pr)}
\end{split}
\end{equation*}
It is important to note that we do not distinguish the sizes nor difficulties of \pullR s.


\section{Method}
Our goal is to determine whether or not our defined measures provide meaningful and interesting insights into the relationship between the OSS Health (\healthAbbv) and Wealth (\wealthAbbv).
We use the following two research questions as a guide into our study:

\begin{itemize}
\item \textbf{$RQ_1$: }\RQone
\item \textbf{$RQ_2$: }\RQtwo
\end{itemize}

To answer both research questions, we conducted an empirical study to measure the Health and Wealth of several OSS projects. 
Similar to Tom Carden's Health and Wealth of Nations, we apply our defined metrics of Health and Wealth.
We use the case study approach to carefully select candidate projects that depict different patterns of Health and Wealth over time. 
To answer $RQ_1$, we analyzed and compared \healthAbbv\ in terms of the labor of both novice and experienced contributors in OSS communities. 
Then finally, to answer $RQ_2$, we investigated the changes of \wealthAbbv\ and corresponding \healthAbbv\ changes.

\section{Studied Projects and Their Community Activities Over Time}
\subsection{Health and Wealth Over Time}
	  \begin{figure*}[!t]
		\center
		\includegraphics[keepaspectratio,scale=0.7]{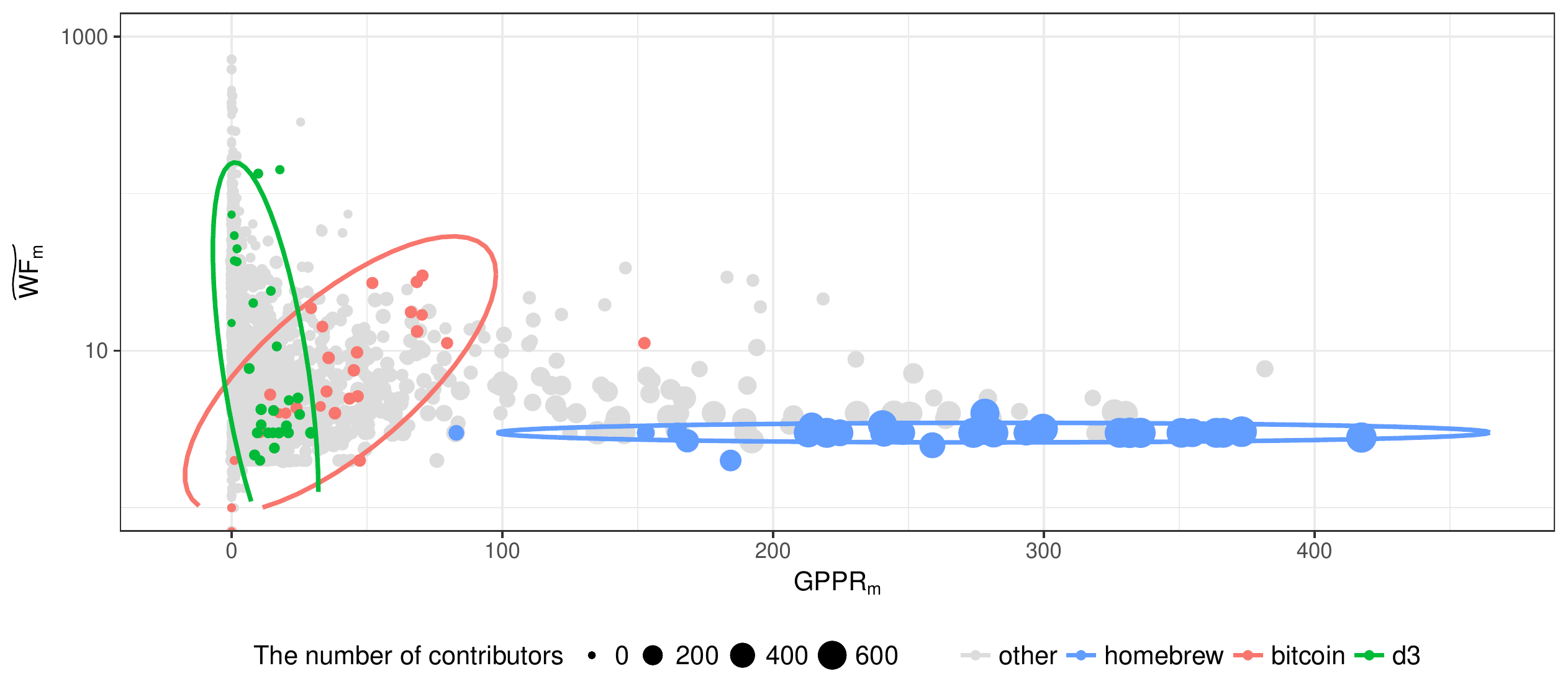}
		 \caption{The Figure shows the $\widetilde{\healthAbbv_m}$ and $\wealthAbbv_m$ of 90 OSS project. from October 2010 to December 2012. In the Figure, we highlight our selected case studies: \typeone~(blue), \typetwo~(red), and \typethree~(green). An interactive demo is available at \url{\movieurl}.}
        \label{fig:health_wealth_all}
      \end{figure*}
\begin{table*}[tb]
\center
  \caption{Summary Statistics of the Selected Case Studies (snapshot as of June 2017)}
  \begin{tabular}{c|c|c|c|c|c}
  \hline\hline
    Project & Life Span & \# of Contributors & \# of Commits & \# of Pull Requests & \# of Issues \\ \hline
    \typethree\ & 6 years and 8 months & 119 & 4,092 & 1,054 & 1,901\\
    \typetwo\ & 7 years and 9 months & 444 & 13,976 & 7,305 & 3,148\\
    \typeone\ & 8 years and 1 month & 5,621 & 63,881 & 33,606 & 17,046\\
    \hline
  \end{tabular}
    \label{tab:project_info}
 \end{table*}
 	\begin{figure*}[p]
		 \center
			 \centering
			\includegraphics[keepaspectratio,scale=0.7]{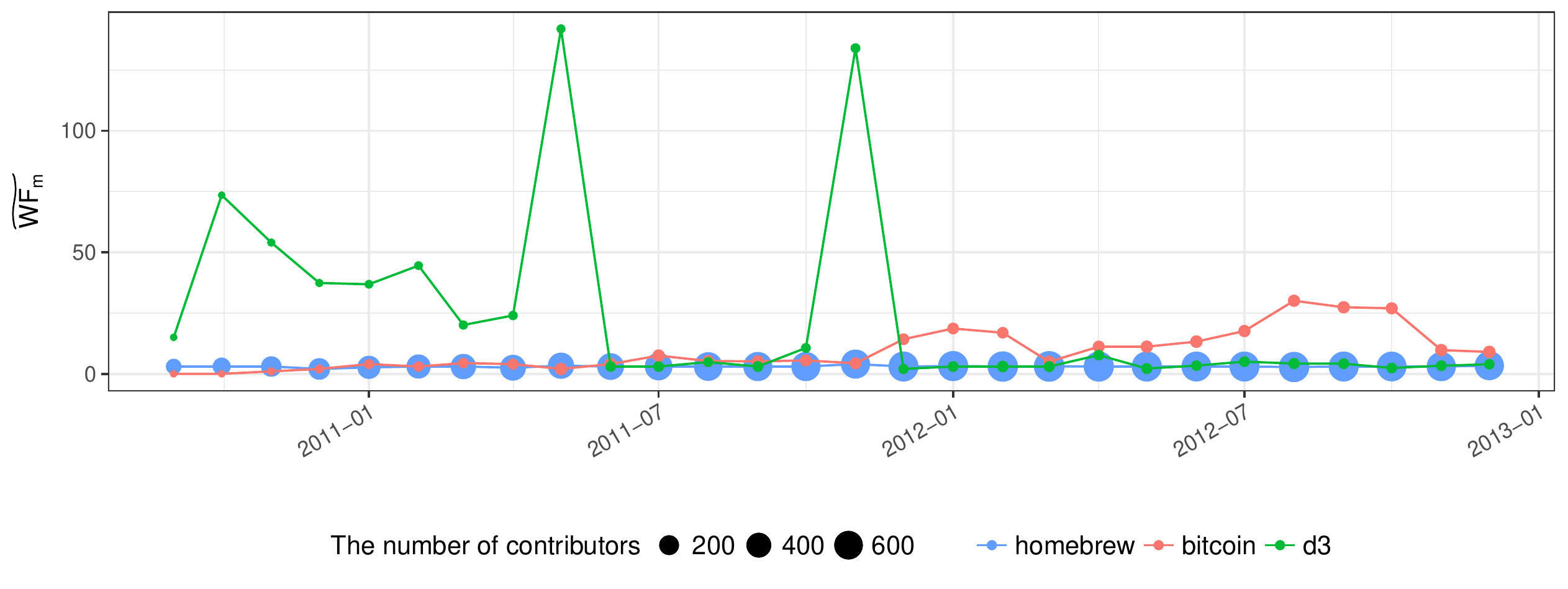}\\
		 \caption{The Figure shows the median values of $\widetilde{\healthAbbv_m}$ from October 2010 to December 2012 for our three case studies.}
		 \label{fig:health}
    \vspace{-3mm}
		 \center
		 \small
		\hspace*{-5mm}
		 \begin{tabular}{cccc}
			\begin{minipage}{0.24\linewidth}
			 \centering
			\includegraphics[keepaspectratio,scale=0.6]{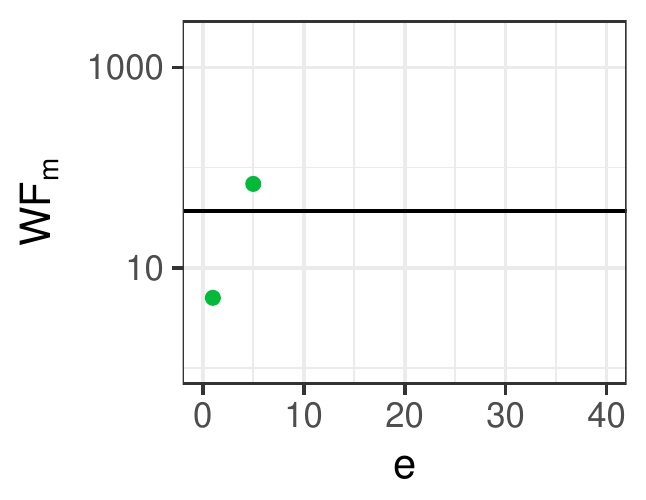}\\
			2011/01 \\
            \vspace{1mm}
            $\widetilde{\healthAbbv_m}$ = 36
			\end{minipage}
			&
			\begin{minipage}{0.24\linewidth}
			 \centering
            \includegraphics[keepaspectratio,scale=0.6]{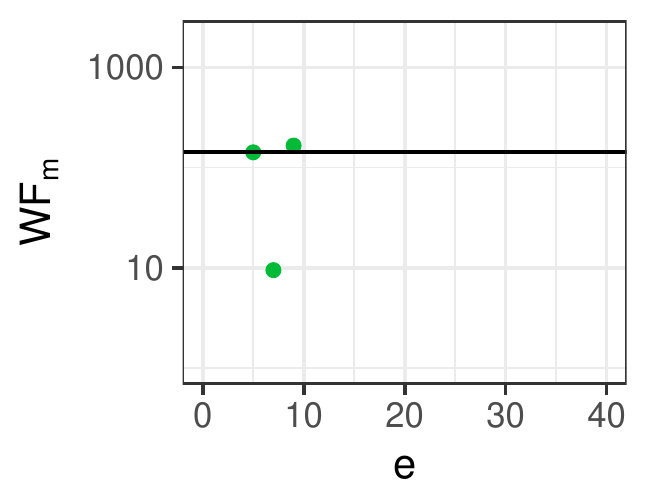}\\
            2011/05 \\
            \vspace{1mm}
			$\widetilde{\healthAbbv_m}$ = 3
            \end{minipage}
			&
			\begin{minipage}{0.24\linewidth}
			 \centering
			\includegraphics[keepaspectratio,scale=0.6]{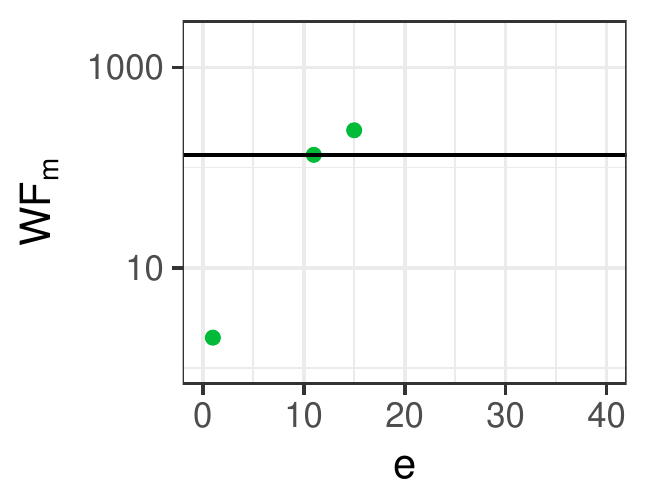}\\
            2011/11 \\
            \vspace{1mm}
			$\widetilde{\healthAbbv_m}$ = 3
            \end{minipage}
			&
			\begin{minipage}{0.24\linewidth}
			 \centering
			\includegraphics[keepaspectratio,scale=0.6]{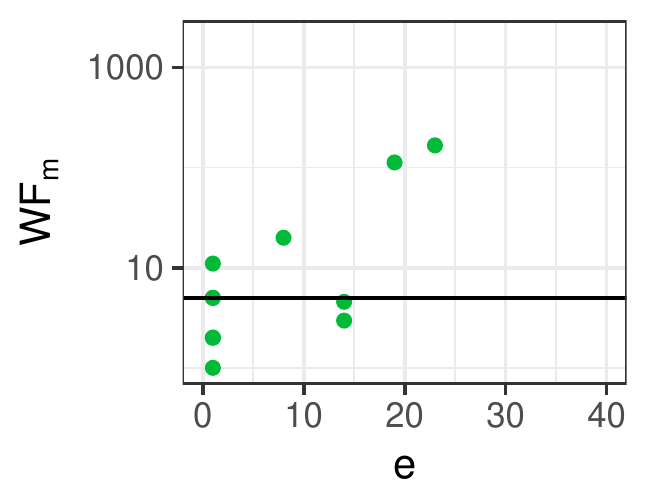}\\
			2012/07\\ 
            \vspace{1mm}
            $\widetilde{\healthAbbv_m}$ = 5
			\end{minipage}
		    \end{tabular}
			\\
            \vspace{2mm}
            (a) These point diagrams show the workforce ($\healthAbbv_m$) and experience (e) for contributors in \typethree. 
        
      
    \vspace{-3mm}
		 \center
		 \small
		\hspace*{-5mm}
		 \begin{tabular}{cccc}
			\begin{minipage}{0.24\linewidth}
			 \centering
			\includegraphics[keepaspectratio,scale=0.6]{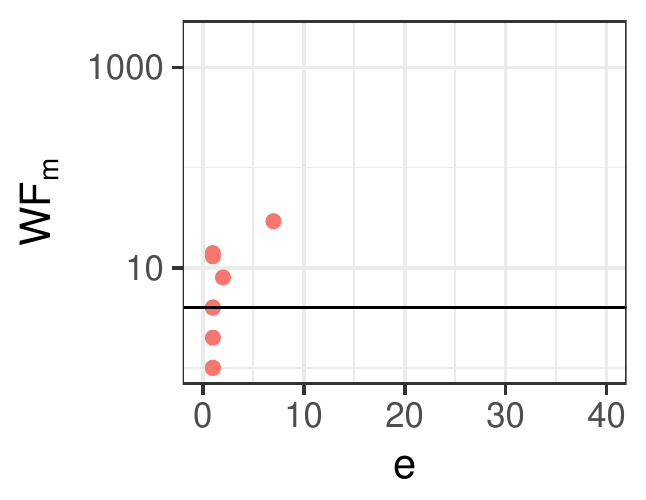}\\
			2011/01 \\
            \vspace{1mm}
            $\widetilde{\healthAbbv_m}$ = 4
			\end{minipage}
			&
			\begin{minipage}{0.24\linewidth}
			 \centering
			\includegraphics[keepaspectratio,scale=0.6]{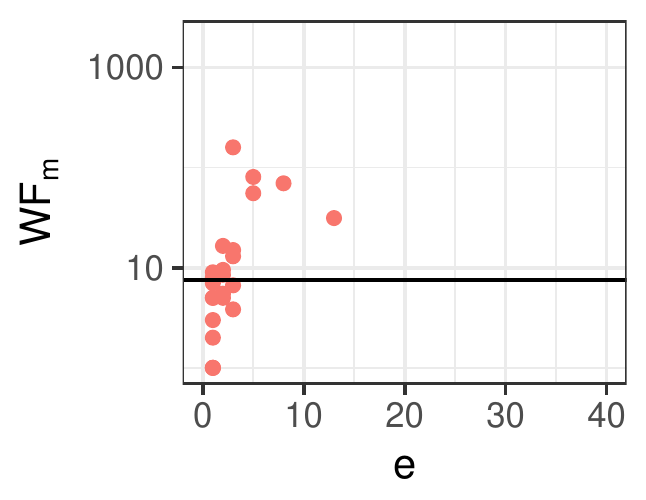}\\
			2011/07 \\
            \vspace{1mm}
            $\widetilde{\healthAbbv_m}$ = 7
			\end{minipage}
			&
			\begin{minipage}{0.24\linewidth}
			 \centering
			\includegraphics[keepaspectratio,scale=0.6]{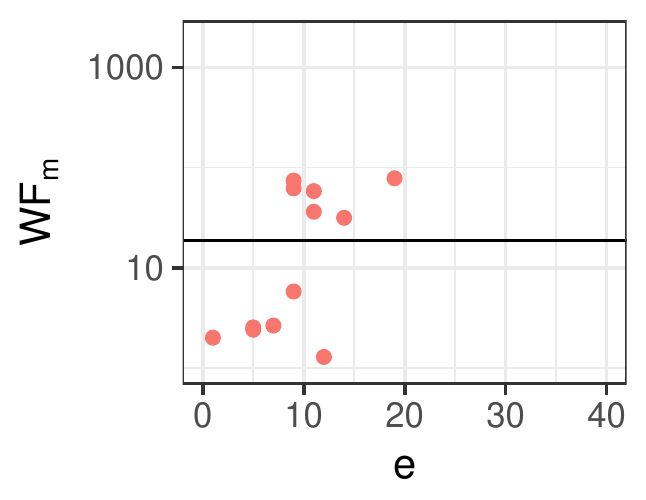}\\
			2012/01 \\
            \vspace{1mm}
            $\widetilde{\healthAbbv_m}$ = 18
			\end{minipage}
			&
			\begin{minipage}{0.24\linewidth}
			 \centering
            \includegraphics[keepaspectratio,scale=0.6]{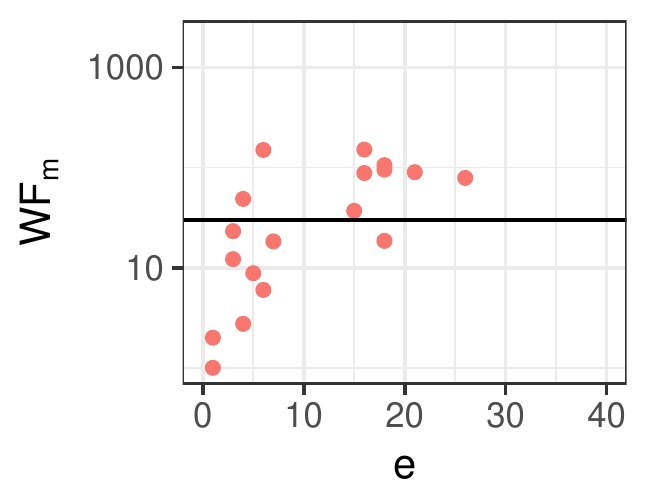}\\
            2012/08\\ 
            \vspace{1mm}
            $\widetilde{\healthAbbv_m}$ = 17
			\end{minipage}
		    \end{tabular}
			\\
            \vspace{2mm}
		 	(b) These point diagrams show the workforce ($\healthAbbv_m$) and experience (e) for contributors in \typetwo

 
		 \center
		 \small
		\hspace*{-5mm}
		 \begin{tabular}{cccc}
			\begin{minipage}{0.24\linewidth}
			 \centering
			\includegraphics[keepaspectratio,scale=0.6]{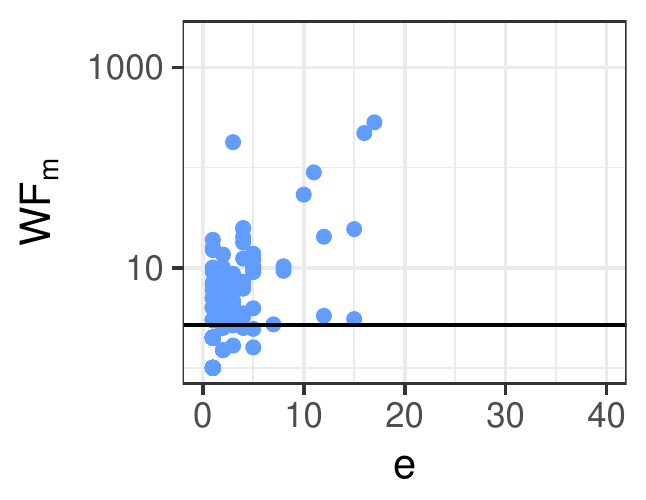}\\
			2011/01 \\
           \vspace{1mm}
           $\widetilde{\healthAbbv_m}$ = 2
			\end{minipage}
			&
			\begin{minipage}{0.24\linewidth}
			 \centering
			\includegraphics[keepaspectratio,scale=0.6]{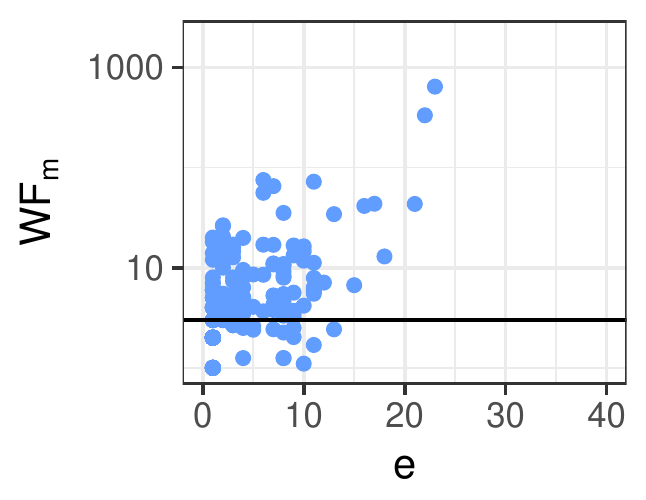}\\
			2011/07 \\
            \vspace{1mm}
            $\widetilde{\healthAbbv_m}$ = 3
			\end{minipage}
			&
			\begin{minipage}{0.24\linewidth}
			 \centering
			\includegraphics[keepaspectratio,scale=0.6]{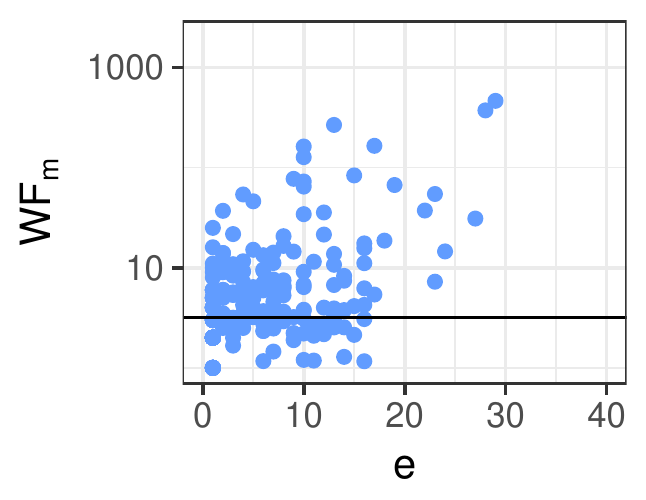}\\
			2012/01 \\
            \vspace{1mm}
            $\widetilde{\healthAbbv_m}$ = 3
			\end{minipage}
			&
			\begin{minipage}{0.24\linewidth}
			 \centering
			\includegraphics[keepaspectratio,scale=0.6]{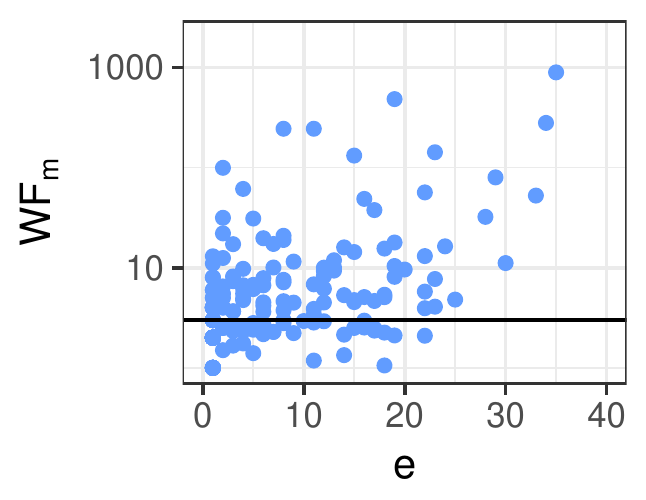}\\
			2012/07\\
            \vspace{1mm}
			$\widetilde{\healthAbbv_m}$ = 3
			\end{minipage}
		    \end{tabular}
			\\
            \vspace{2mm}
            (c) These point diagrams show the workforce ($\healthAbbv_m$) and experience (e) for contributors in \typeone
		 \caption{The Point diagram of $\healthAbbv_m$ and activity periods of all contributors. Horizontal line shows median of $\healthAbbv_m$.}
		 \label{fig:point_diagram}
	\end{figure*}
 	\begin{figure*}[!t]
		 \center
			 \centering
			\includegraphics[keepaspectratio,scale=0.7]{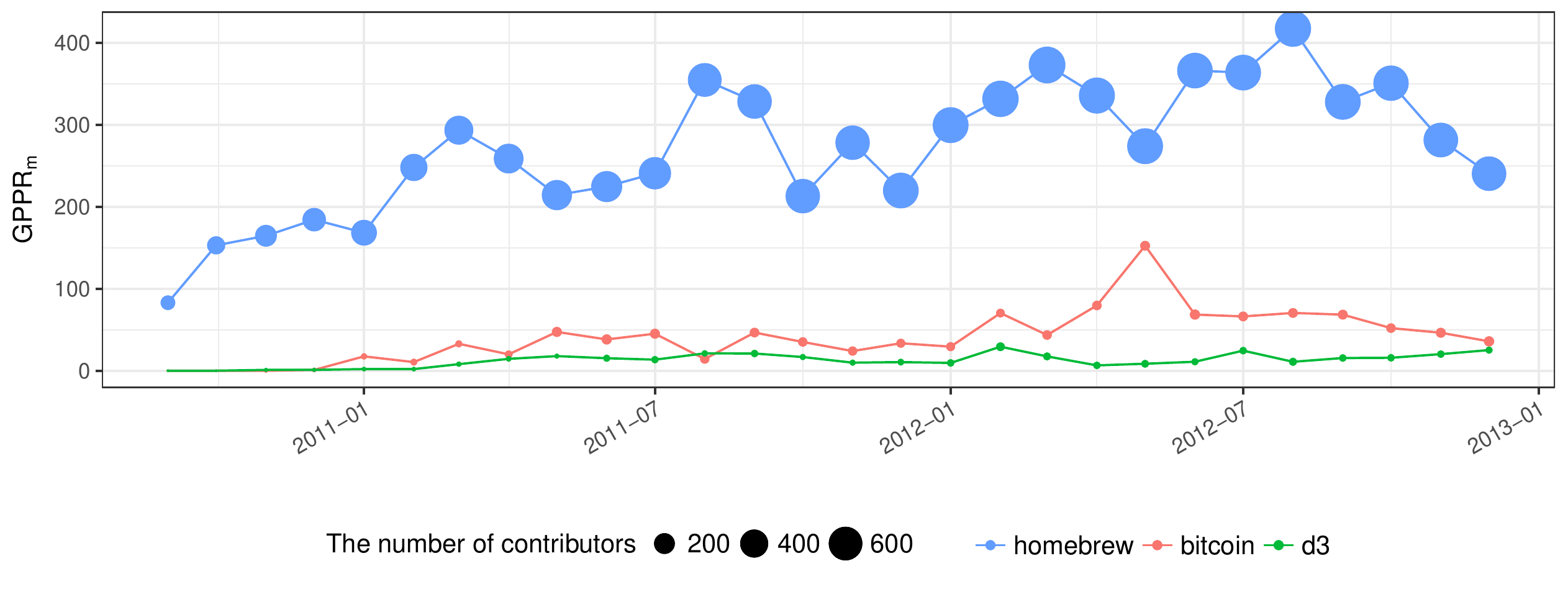}\\
		\caption{The $\wealthAbbv_m$ from October 2010 to December 2012 in for our three case studies. Note that the size each point is indicative of the total number of contributors for that project.}
		 \label{fig:wealth}
		\end{figure*}

For our empirical study, we first collected and analyzed 90 OSS projects provided by the GHTorrent \cite{Gousios2013}.
From this dataset, we were able to select and gather two of the related contribution activities (i.e., \texttt{commits} and \texttt{pull requests}) for the \healthAbbv~and \wealthAbbv~metric calculations.

Figure \ref{fig:health_wealth_all} shows the Health and Wealth of all projects tracked from October 2010 to December 2012.
We highlight three projects that depict different patterns in the relation of Health and Wealth. 

From the figure, we were able to identify three types of OSS evolution of the Health and Wealth:

\begin{itemize}
\item \textit{Consistent Wealth but changes in healthiness:} These OSS projects show consistent Wealth, however, these projects experience changes in its Health.
\item \textit{Changes in both Health and Wealth rates:}
These projects experience changes in both its Health and Wealth over time.
\item \textit{Changes in wealthiness while keeping consistent Health:}
These OSS projects depict Health at a consistent rate, however, the projects increase their Wealth.
\end{itemize}



\subsection{Case Study Selections}

%
Figure \ref{fig:health_wealth_all} shows three OSS projects that depict three distinct patterns between the Health and Wealth of an OSS project over time.
Table \ref{tab:project_info} shows the detailed information of each of these projects.
\begin{itemize}
\item \textbf{\typethree} (green color) \footnote{\typethree: https://github.com/d3/d3} is a JavaScript library for visualizing data using web standards. 
As shown in Table \ref{tab:project_info} this project is the youngest and smallest of the three selected projects, having the smallest contributors, commits, pull requests and issues. 
We can see that \typethree\ had experienced \textit{changes in \healthAbbv~Health, while keeping a low but consistent \wealthAbbv~Wealth}.
\item \textbf{\typetwo} (red color)\footnote{\typetwo: https://github.com/bitcoin/bitcoin} is software that enables the use of currency referred to as bitcoin.
As shown Table \ref{tab:project_info}, this project is the middle of the three selected projects as for all items. 
Out of the three selected projects, we find that \typetwo~exibits the most number of commits per one contributor.
We can see that \typetwo\ had experienced \textit{changes in both \wealthAbbv~wealth and \healthAbbv~Health}.
\item \textbf{\typeone} (blue color) \footnote{\typeone: https://github.com/Homebrew/brew} is a software package management system that simplifies the installation of software on Apple's Mac OS operating system.
As shown Table \ref{tab:project_info}, this project is the oldest and has the biggest statistics of the three selected project (i.e., has the most community of contributors, commits, pull requests and issues). 
We can that \typeone\ had experienced \textit{increase in \wealthAbbv~wealth, while keeping a consistently low \healthAbbv~Health}.
\end{itemize}

It is difficult to judge whether or not these selected projects are successful.
However, considering that since these projects have been active for 6 to 8 years, we assume that they are representatives of typical OSS projects and their communities.

\section{Results}
\subsection*{\textbf{$RQ_1$: \RQone}}
Figure \ref{fig:health} shows the median \healthAbbv\ for our three case studies over time.
For a deeper analysis of Health in terms of \healthAbbv\ and experience levels of contributors, Figure \ref{fig:point_diagram} depicts four snapshots of \healthAbbv\ along with contributor experiences.
From these Figures, we are able to make three observations:

\vspace{2mm}
\begin{quote}
\textit{``Less wealthy projects may require additional efforts (i.e., higher \healthAbbv ) from their more experienced contributors.''}
\end{quote}

Figure \ref{fig:health} shows that overall, less Wealth projects (green) occasionally experience bursts (i.e., depicted by the spikes in \healthAbbv\ ) from its community workforce.
Furthermore, we find in Figure \ref{fig:point_diagram}(a) and Figure \ref{fig:point_diagram}(b) that \typethree\ project experienced more \healthAbbv\ from their more experienced contributors.
We conjecture that less wealthy projects (i.e., \typethree\ and \typetwo ) occasionally require more \healthAbbv\ from their experienced contributors.

\vspace{2mm}
\begin{quote}
\textit{``Wealthy projects attract and rely on the inexperienced casual workforce.''}
\end{quote}

Figure \ref{fig:health} shows that \typeone\ keeps a consistent low \healthAbbv\.
One explanation, for the low \healthAbbv\, may be accounted by the high number of contributors of less experienced (i.e., casual contributors) in the community.  
We conjecture that the key to \typeone 's Wealth is the ability to attract and keep these casual contributors.


\vspace{2mm}
\begin{quote}
\textit{``Experienced contributors are major workforce of an OSS project.''}
\end{quote}

Generally, from Figure \ref{fig:point_diagram}, we observe that the highest rates of \healthAbbv\ is from the more experienced contributors. 
This results confirms the common notions that experienced contributors are indeed major workforce of an OSS project.


In summary, to answer $RQ_1$, we find that \textit{ wealthy projects may not necessarily depend only on higher workforce, as they can rely on many contributions from their casual contributors. Consequently, less wealthy project may require additional efforts from their more experienced workforce}.

\subsection*{\textbf{$RQ_2$: \RQtwo}}
 
Figure \ref{fig:wealth} shows the \wealthAbbv\ for our three case studies over time. 
Using this Figure, we make the following observation:

\vspace{2mm}
\begin{quote}
\textit{``Wealthy projects do not depend only on higher workforce.''}
\end{quote}

In Figure \ref{fig:wealth}, we can clearly observe the difference in \wealthAbbv s between the wealthiest project (i.e., \typeone ) and less wealthy projects (i.e., \typetwo\ and \typethree ).

It is interesting to note that even though \typeone\ experiences a constant Wealth of pull requests.
Under deeper manual analysis, we found that \typeone\ contributors not to be as effective, with many contributors ignoring a significant amount of pull requests.
Therefore, corresponding with $RQ_1$, we conclude that such projects rely the size of its casual contributors to maintain its Wealth.


On the other hand, less wealthy projects remain with a lower \wealthAbbv s, requiring more Health \healthAbbv\ from its contributors.
However, there exists cases when a project may increase its Wealth.
For example, \typetwo\ occasionally increased its \wealthAbbv s.
A manual analysis revealed that a single contributor had submitted all 58 {\PR} in a short period, thus increasing the project's Wealth.
In fact, their contribution accounts for about one-third the total \wealthAbbv\ (177 \PR) during this period.
We conjecture that such experienced workforce does have an influence on Wealth.
%

In summary, to answer $RQ_2$, we find that \textit{less wealthy projects rely mainly on their active workforce to maintain their Wealth or increase their Wealth.}

\section{Related Work}

Ye et al examined the structure of Free and Open Source Software (F/OSS) communities and the co-evolution of F/OSS systems and communities.
They report F/OSS systems and communities generally co-evolve, they co-evolve differently depending on the goal of the system and the structure of the community\cite{ye2004}. 
Our study also mentions product evolution and community activities through the analyzing the Health and Wealth in OSS projects. 
This study can help to understand the co-evolution of OSS systems and communities.

Gousios et al. explored how pull-based software development works in OSS projects\cite{Gousios2014,Gousios2015,Gousios2016}. 
They find that the pull request model offers fast turnaround, increased opportunities for community engagement and decreased time to incorporate contributions. 
Also, our study presents a measure which is by number of Pull Requests as \wealthAbbv.
This measure is very useful to understand whether a project can takes advantage of pull requests.

Zhou et al. studied long-term contributors (LTC), analyzing the behavior of individual participants in
Gnome and Mozilla \cite{Zhou2012}.
They report that future LTCs tend to be more active and show more community-oriented attitudes than do other joiners during their first month.
Also, Pinto et al. analyzed about activities of casual contributors\cite{Pinto2016}.
Casual contributors are that developers do not want to become active members.
They describe casual contributors that foster diversity and collaboration.　
Our study presents a measure which is contributors' activities as \healthAbbv. 
We think this measure is very important to clarify whether a project has experienced contributors and casual contributors.

\section{Conclusions}
Economist consider that Wealth and Health is very important to clarify the condition, features for future of a country.
In this study, we propose the Health and Wealth in the context of OSS projects. 
We focus on the number of submitted and closed pull requests and experiences of contributors activity, and define two measures of \health\ (\healthAbbv) and \wealth\ (\wealthAbbv). 

From these measures, we identified three situations of OSS evolution of the Health and Wealth. 
First, we analyzed projects that have changes in healthiness while keeping consistent Wealth.
Second, we then studied changes in wealthiness while keeping consistent Health.
Finally, we studied the changes in both Health and Wealth rates.
From this analysis, we find that wealthy projects attract and rely on the inexperienced casual workforce, while less wealthy projects may require additional efforts from their more experienced contributors.

Our future work includes a more in depth study with metrics adapted from other fields, and bigger dataset to clarify the relationship between the Wealth and Health of OSS projects, and the Wealth as the economic and the Health as a demography.

\end{document}